\documentclass[12pt]{article}
\usepackage{amsmath}
\usepackage{mathrsfs}
\usepackage{amssymb}
\usepackage{bm}
\begin{document}
\begin{titlepage}
%\begin{flushright}
%v3.10,\\
%\today
%\end{flushright}

\begin{center}
\begin{Large}
{\bf  Nambu Quantum Mechanics \\ on Discrete 3-Tori}
\end{Large}

\vskip1.0truecm

M. Axenides$^{(a)}$\footnote{E-Mail: axenides@inp.demokritos.gr},
E. G. Floratos$^{(b)}$\footnote{E-Mail: mflorato@physics.uoa.gr}
and S. Nicolis$^{(c)}$\footnote{E-Mail: Stam.Nicolis@lmpt.univ-tours.fr}

\vskip1.0truecm

{\sl
$^{(a)}$~National Research Center ``Demokritos''\\
  15310 Aghia Paraskevi, Athens, Greece\\

\vskip0.5truecm

$^{(b)}$~Nuclear and Particle Physics Section, University of Athens\\
15771 Athens, Greece\\

\vskip0.5truecm

$^{(c)}$~CNRS--Laboratoire de Math\'ematiques et Physique Th\'eorique (UMR 6083)\\
F\'ed\'eration D\'enis Poisson (FR 9164)\\
Universit\'e de Tours ``Fran\c{c}ois Rabelais'', Parc Grandmont,  37200 Tours,
France
}
\end{center}

\begin{abstract}

We propose a quantization of linear, volume preserving ,maps on the discrete
and finite 3-torus $\mathbb{T}_N^3$ represented by elements of the group
$SL(3,\mathbb{Z}_N)$. These flows can be considered as special motions of the
Nambu dynamics (linear Nambu flows) in the three dimensional toroidal
phase space and are characterized by invariant vectors $ \bm{a} $ of $\mathbb{T}_N^3$.
We quantize all such flows which are necessarily restricted
on a planar two-dimensional  phase space, embedded in the 3-torus,
transverse to the vector $\bm{a}$ . The corresponding maps belong to the little group
of $\bm{a} \in SL(3,\mathbb{Z}_N) $ which is an $SL(2,\mathbb{Z}_N)$ subgroup.
The associated linear Nambu maps are generated
by a pair of linear and quadratic Hamiltonians (Clebsch-Monge potentials of the flow)
 and the corresponding quantum maps, realize the metaplectic representation of  $SL(3,\mathbb{Z}_N)$ on the discrete group of
three dimensional magnetic translations i.e.  the non-commutative 3-torus with
deformation parameter the $N-$th root of unity. 

Other potential applications of our construction are related to the quantization of deterministic chaos in turbulent maps as well as
to quantum tomography of three dimensional objects.

\end{abstract}
\end{titlepage}

\section{Introduction}

Recently, due to the progress in understanding  the dynamics of the low energy
effective field theories for systems of multiple membranes,
in analogy with the AdS/CFT correspondence~\cite{bagger_lambert},
new algebraic structures,  the metric 3-algebras,  which are related to the
quantization problem of Nambu 3-brackets~\cite{nambu,tak,hop,yoneya,minic,kawamura}
 have attracted considerable interest
~\cite{zachos,gustavsson,papadopoulos,matsuo_ho,matsuo_ho2,axenides_floratos}.

In \cite{axenides_floratos} we discussed in detail the relation between metric 3-algebras and
Nambu 3-brackets and we proposed a specific quantization method inspired by
the work of Takhtajan~\cite{tak}.

Nambu in his classic paper~\cite{nambu} introduced a generalization of classical mechanics, where the role of canonical transformations
of Hamiltonian mechanics is played by the general volume preserving
diffeomorphism group of a  manifold of any dimension, considered as the
corresponding phase space.
For example, in three dimensional Euclidean space, we consider incompressible flows and the particle trajectory,flow equations,are the Nambu
dynamical equations,  first order in time, differential
equations that generalize Hamilton's equations of motion.

In the next section we recall that,  in this case,
there are two Hamiltonians $H_1,H_2$ corresponding to the Clebsch-Monge
potentials of the flow ,which are conserved and their constant values define
 a double family of intersecting surfaces ,the intersections of which define the trajectories of the (test)
particles, carried by the flow.
The interpretation we adopted in \cite{axenides_floratos}, is that the surfaces defined by the
second potential are   conventional, two-dimensional phase spaces, foliating the 3-space and the first
potential defines  conventional Hamiltonian mechanics on these phase spaces.
There is  a common  induced Poisson bracket on the two-dimensional phase spaces
embedded in $\mathbb{R}^3$ and the Nambu equations on these phase spaces take the standard Hamiltonian form. 

The quantum mechanics of this system, must follow the quantization
of the induced Poisson structure, which depends on the symplectic structure on the $H_2$ surfaces.
If the $H_2$ Hamiltonian is linear and $H_1$ is quadratic we have an
incompressible linear flow in $\mathbb{R}^3$.

 In the present work we consider the classical discretization of these
linear flows(maps), in toroidal discrete three dimensional phase space and their quantization.
The corresponding quantum 3-dimensional phase space, is a  non-commutative
three-torus with rational values of the noncommutativity
parameter~\cite{rieffel}. 
The classical linear maps of $SL(3,\mathbb{Z}_N)$ are related to strong 
arithmetic(deterministic) chaos~\cite{vivaldi}
and can be considered as  discrete toy models for turbulence 
on $\mathbb{T}^3$~\cite{lima}.
These considerations and the possible physical inerpretation 
of their quantized version will be discussed elsewhere.

The plan of the paper is as follows:

In sec.~\ref{Nambu} we recall  the formulation of Nambu dynamics in
$\mathbb{R}^3$ and $\mathbb{T}^3$, in sec.~\ref{LNF_sec}  we  consider the deterministic, chaotic, linear maps,
analogs of the Arnold cat maps (but in three dimensions) which are elements of
$SL(3,\mathbb{R})$ and introduce their Lie algebra. In sec.~\ref{discrete_phsp} we pass to
the discretized 3-torus and consider corresponding maps, which are elements of
$SL(3,\mathbb{Z}_N)$ and, indeed, belong to the little group,
$SL(2,\mathbb{Z}_N)$, of the invariant vectors normal to the planes of the flow.
In sec.~\ref{nc_3T}  we consider the non-commutative,rational 3-torus and we
construct the Heisenberg--Weyl group, that corresponds to the linear Nambu
flows as well as the associative
quantum 3-algebra for the foliation of the 3-torus by the normal vectors .

In sec.~\ref{Quant3T} we present the quantization of these maps.  The quantum
maps (realized by unitary $N\times N$  matrices)are constructed explicitly by
imposing  the  (metaplectic) representation of $SL(3,\mathbb{Z}_N)$, induced
by the little group, $SL(2,\mathbb{Z}_N)$

We end with our conclusions, interpretation of our results and discuss some emergent applications.

\section{Nambu Mechanics in $\mathbb{R}^3$ and $\mathbb{T}^3$}\label{Nambu}
 In his classic paper~\cite{nambu} Y. Nambu generalized classical
Hamilton--Poisson mechanics by considering arbitrary dimensions for the phase
space, replacing the canonical transformation symmetry by the volume
preserving diffeomorphisms~\cite{zachos}. In the particular case
of three dimensional, flat,
phase space, one needs {\em two} "Hamiltonian" functions and the Nambu equations of
motion take the form
\begin{equation}
\label{nambu_eq}
\frac{dx^i}{dt} = \left\{x^i, H_1, H_2\right\}
\end{equation}
$i=1,2,3$ with initial conditions $x^i(t=0)\equiv x^i(0)$.

 The 3-bracket for functions, $f,g,h\in {\mathscr C}^\infty(\mathbb{R}^3)$ is defined as
\begin{equation}
\label{3-bracket}
\left\{ f,g,h\right\} \equiv \varepsilon^{ijk}\partial_i f\partial_jg \partial_k h
\end{equation}
The Nambu bracket is invariant under general coordinate transformations, which
preserve the volume, i.e. for $y^i  = u^i(\bm{x})$ with $J=\mathrm{det}\,\partial
u^i/\partial x^j = 1$. These transformations form the volume preserving
diffeomorphism group, $\mathrm{SDiff}(\mathbb{R}^3)$ under the composition of
mappings~\cite{marsden_weinstein}.
For infinitesimal transformations,
\begin{equation}
\label{voldiff_infini}
x^i\to x^i+v^i(\bm{x})
\end{equation}
with $v^i(\bm{x})$ a divergenceless vector field. This transformation 
 defines the flow
\begin{equation}
\label{flow}
\frac{d\bm{x}}{dt} = \bm{v}(\bm{x})
\end{equation}
with 
 corresponding generators, $X(\bm{v})\equiv -v^i\partial^i$.They  form the Lie
algebra
\begin{equation}
\label{Lie_algebra}
\left[X(\bm{v}), X(\bm{u})\right] = X(\bm{w})
\end{equation}
where
$$
w^i\equiv \varepsilon^{ijk}\partial^j( \left(\bm{v}\times\bm{u}\right)^k)
$$
The right-hand-side  of eq.~(\ref{flow}) can be written in terms of two Clebsch-Monge
potentials \cite{lamb}, $H_1$ and $H_2$
\begin{equation}
\label{solution}
v^i = \varepsilon^{ijk}\partial^jH_1\partial^k H_2
\end{equation}
The Lie algebra of $\mathrm{SDiff}(\mathbb{R}^3)$, in the Clebsch-Monge gauge,
becomes
\begin{equation}
\label{clebsch_gauge}
\left[X(H_1, H_2), X(H_3, H_4)\right] =
X( X(H_1,H_2)H_3, H_4 ) +
X( H_3, X(H_1,H_2)H_4 )
\end{equation}
Notice that $X(H_1, H_2)H_3=-\left\{H_1, H_2, H_3\right\}$, i.e. they
realize the Nambu bracket.

From the above we see that the Nambu equations of motion describe
incompressible flows in $\mathbb{R}^3$ and the solutions represent
the integral curves of the flow $v^i(H_1, H_2)$.

The 3-bracket has certain interesting properties~\cite{tak}:
It
\begin{itemize}
\item is {\em multilinear} in $f,g,h$
\item is {\em antisymmetric} in $f,g,h$
\item has the {\em Leibniz} property:
\begin{equation}
\label{leibniz}
\left\{f_1f_2, g,h\right\} = f_1\left\{f_2,g,h\right\} +
\left\{f_1,g,h\right\}f_2
\end{equation}
\item Satisfies the  {\em  fundamental identity}
\begin{equation}
\label{fund_ident}
\left\{ \left\{ f_1, f_2, f_3\right\}, f_4, f_5\right\} +
\left\{  f_1, \left\{f_4, f_2,f_3\right\}, f_5\right\} +
\left\{f_1, f_4, \left\{ f_5, f_2,f_3\right\}\right\} =
\left\{ \left\{ f_1, f_4, f_5\right\}, f_2, f_3\right\}
\end{equation}
\end{itemize}
The fundamental identity can be proved by applying both sides of
eq.~(\ref{clebsch_gauge}) with $ f_{i}=H_{i}, i=1,\cdots,4 $, on a function $f_{5}$,
where
$f_i\in\mathscr{C}^\infty(\mathbb{R}^3), i=1,\cdots,5 $.

We can obtain Liouville's eq. for an aribitrary observable,
that does not depend explicitly on time,
\begin{equation}
\label{liouville}
\frac{df}{dt} = \left\{ f, H_1, H_2 \right\}
\end{equation}
This equation implies  the conservation of the
"Hamiltonians" $H_1$ and $H_2$ under the flow and, thus, the particle's
trajectory lies on the intersection of the two surfaces in
$\mathbb{R}^3$,defined by $H_1$ and $H_2$, given the initial conditions,
$\bm{x}(0)$.
Its formal solution may be written as
\begin{equation}
\label{sol_liouv}
f(\bm{x}) = e^{-tX(H_1, H_2)}f(\bm{x}(0))
\end{equation}
We shall need later the toroidal compactification of $\mathbb{R}^3$,
$\mathbb{T}^3\equiv \mathbb{R}^3/\mathbb{Z}^3$. The smooth functions,
$f\in\mathscr{C}^\infty(\mathbb{T}^3)$ may be written as
$$
f(\bm{x}) = \frac{1}{(2\pi)^{3/2}}
\sum_{\bm{n}\in\mathbb{Z}^3}f_{\bm{n}}\cdot e^{\mathrm{i}\bm{n}\cdot\bm{x}}
$$
The Poisson bracket of two functions $f, g$ on the 3-torus is defined as
\begin{equation}
\label{PoissonB_3T}
\left\{f, g\right\}\equiv \varepsilon^{ijk} a^i \partial^j f\partial^k g
\end{equation}
once a constant vector
$ \bm{a} \in \mathbb{R}^{3} $ is given and the corresponding Poisson manifold is denoted
with $ \mathbb{T}^{3}_{\bm{a}}$.
The Poisson algebra on the basis , $e^{\mathrm{i}\bm{n}\cdot\bm{x}}, 
\bm{n} \in \mathbb{Z}^{3}$ is
given by
\begin{equation}
\label{PoissonAlg3T}
\left\{
e^{\mathrm{i}\bm{n}\cdot\bm{x}},e^{\mathrm{i}\bm{m}\cdot\bm{x}}\right\} =
-\mathrm{det}(\bm{a},\bm{m},\bm{n})e^{\mathrm{i}(\bm{m}+\bm{n})\cdot\bm{x}}
\end{equation}
The corresponding algebra, $\mathrm{SDiff}(\mathbb{T}^3)$ can be expressed in
terms of the Nambu bracket \cite{hop,matsuo_ho,axenides_floratos}
\begin{equation}
\label{torus3}
\left\{
e^{\mathrm{i}\bm{n}_1\cdot\bm{x}},
e^{\mathrm{i}\bm{n}_2\cdot\bm{x}},
e^{\mathrm{i}\bm{n}_3\cdot\bm{x}}
\right\} =
-\mathrm{i}\times \mathrm{det}(\bm{n}_1,\bm{n}_2,\bm{n}_3) e^{\mathrm{i}(\bm{n}_1+\bm{n}_2+\bm{n}_3)\cdot\bm{x}}
\end{equation}

 The compactification on the torus or on the sphere can also be considered
  as an infra-red cutoff for growing modes of incompressible flows over large
  distances in fluid dynamics~\cite{Moff}
%--------------------------------------------------------------------------------------------------------
%--------------------------------------------------------------------------------------------------------
\section{Linear Nambu Flows}\label{LNF_sec}

%--------------------------------------------------------------------------------------------------------%--------------------------------------------------------------------------------------------------------

In this work we focus on the case of {\em linear} Nambu flows, which can be
derived from a pair of Hamiltonians, $H_2  = \bm{a}\cdot\bm{x}$ and
$H_1 = (1/2)(\bm{x},{\sf B}\bm{x})$, where $\bm{a},\bm{x}\in\mathbb{R}^3$ and
${\sf B}$ is a real, symmetric, $3\times 3$ matrix. The corresponding
trajectory of the linear Nambu flow (LNF) is given by
\begin{equation}
\label{nambu_form}
\frac{dx^i}{dt} = \varepsilon^{ijk}a^jB^{kl}x^l\equiv x^l{\sf M}^{li}
\end{equation}
The solutions, given an initial condition, $x^i(0)$, lie on the intersection
of the plane with normal vector, $\bm{a}$ and the quadratic surface given by
$H_1=(1/2)(\bm{x}(0),{\sf B}\bm{x}(0))$. We can integrate the equation of
motion explicitly and find
\begin{equation}
\label{LNF}
\bm{x}(t) = \bm{x}(0)e^{t{\sf M}}
\end{equation}
Since the matrix ${\sf M}$ is traceless, ${\sf A}\equiv e^{{\sf M}}$
is an element of the
group $SL(3,\mathbb{R})$. The converse is {\em not} true, i.e. every traceless
matrix ${\sf M}$ (element of the Lie algebra, $sl(3,\mathbb{R})$) defines an
incompressible flow, which, in general, does not admit a representation in
terms of a linear and a quadratic Hamiltonian pair.
If we require, in addition, that ${\sf M}$ have an eigenvector with zero
eigenvalue, then it can be shown that it is of the ``Nambu form'',
eq.~(\ref{nambu_form}).

It is possible to compactify the LNF on ${\mathbb T}^3$, if we consider the
linear evolution equation, eq.~(\ref{nambu_form}), modulo the size of the
torus, i.e. we take $x^i$, $i=1,2,3$, to belong to the elementary cell,
$x^i\equiv x^i + L^i$, where $L^i$ is the length of the torus along direction $x^i$. We
choose our units so that $L^i=2\pi$. The action of the
matrix ${\sf A}$ on every
point of $\mathbb{T}^3$ is then
taken modulo $2\pi$. These flows are known in the
literature as {\em toral automorphisms}~\cite{arnold}.
The motion in this case, even though the
equation is linear, can be chaotic, depending on the matrix elements of ${\sf
  A}$. We can check that, for LNF in $\mathbb{R}^3$, we have, essentially,
a reduction to a two-dimensional phase space problem on the plane orthogonal
to the vector $\bm{a}$. In the case of $\mathbb{T}^3$, if the vector has rational
 components, then we have a finite number of different images of the
plane; if, however, the components are irrationals, then we have a truly
three-dimensional evolution for the system.

Considering the algebra of all LNF, we characterize the corresponding
generators   by a vector
$\bm{a}\in\mathbb{R}^3$ and a symmetric, $3\times 3$  matrix ${\sf B}$
\begin{equation}
\label{generator}
X(\bm{a}, {\sf B}) = -\varepsilon^{ijk}a^jB^{kl}x^l\partial^i
\end{equation}
Their Lie algebra closes as follows
\begin{equation}
\begin{array}{l}
\displaystyle
\left[ X(\bm{a}_1, {\sf B}_1), X(\bm{a}_2, {\sf B}_2)\right] =
X(\bm{a}_3, {\sf B}_2) + X(\bm{a}_2,{\sf B}_3)\\
\\
\displaystyle
a_3^l = \varepsilon^{ijk}a_2^ia_1^jB_1^{kl}\\
\\
\displaystyle
B_3^{lm} = 2\varepsilon^{ijk}a_1^jB_1^{kl}B_2^{im}
\end{array}
\end{equation}
Since this algebra contains a total of eight independent parameters, it can
generate $SL(3,\mathbb{R})$, i.e. consecutive application of different LNF
give rise to an $SL(3,\mathbb{R})$ flow, which is not, necessarily, LNF.

%--------------------------------------------------------------------------------------------------------%--------------------------------------------------------------------------------------------------------

\section{The Discrete Phase Space of Linear Nambu Flows}\label{discrete_phsp}

%--------------------------------------------------------------------------------------------------------%--------------------------------------------------------------------------------------------------------
The simplest discretization of
 $\mathbb{T}^3_{\theta}$ (where $\bm{\theta}\in\mathbb{R}^3$) can be constructed by considering only points with
rational coordinates, $x^i = 2\pi k^i/N$, $k^1,k^2,k^3$ \mbox{integers modulo} $N$,
whose denominator is a fixed prime number,
$N$. Discretization of flows is necessary in order to provide a ultraviolet cutoff to
non-linear, classical, instabilities~\cite{rajeev}.
 This set forms a three-dimensional, abelian, group, $\mathbb{T}_N^3$
 under addition of coordinates modulo $2\pi$. The linear maps, which define
the evolution in this discrete phase space, are elements of
 $Sl(3,\mathbb{Z}_N)$, i.e. $3\times 3$ integer matrices with entries taken
 modulo $N$ and determinant equal to one (modulo $N$). The discrete time
  evolution, for any ${\sf A}\in SL(3,\mathbb{Z}_N)$,
 is given as
\begin{equation}
\label{classical_evol}
\bm{x}_{n+1} = \bm{x}_n\cdot {\sf A}
\end{equation}
whose solution is
\begin{equation}
\bm{x}_n = \bm{x}_0\cdot {\sf A}^n,\,\,n=0,1,2,\ldots
\end{equation}
Since the group $SL(3,\mathbb{Z}_N)$ is finite, all orbits are periodic and
there exist interesting special motions, which form subgroups thereof, namely,
shears, rotations and dilatations (cf.also below). The shears form the 
 discrete Heisenberg-Weyl subgroup $ HW_{N} $, which  is the set of 
elements ( acting
on the right side of points of $\mathbb{T}_{N}^{3}$).
\begin{equation}
\label{HW}
{\sf T}(a,b,c) = \left( \begin{array}{ccc}1&0&0\\a&1&0\\c&b&1 \end{array}\right)
\ \ \ \ \ \ \ \ \ \ \
a,b,c \in\mathbb{Z}_N
\end{equation}
\begin{equation}
{\sf T}( a_{1}, b_{1}, c_{1}) \ {\sf T}( a_{2}, b_{2}, c_{2} ) = {\sf T}( a_{1} + a_{2}, b_{1} + b_{2},
c_{1} + c_{2} + b_{1} a_{2} ) \ \ \ \ \ \ \
a_{i} , b_{i} , c_{i}\in\mathbb{Z}_N\ \ i=1,2
\end{equation}
an inverse element
\begin{equation}
 {\sf T}^{-1}(a,b,c)= {\sf T}(-a,-b,-c + ab)
\end{equation}
and center ( equal to $Z_{N}$) generated by the element $\Omega = T(0,0,1)$.
The commutation
relations of two elements are given by :
\begin{equation}
{\sf T}(a_{1},b_{1},c_{1}) {\sf T}(a_{2},b_{2},c_{2}) \ = \ \Omega^{b_{1}a_{2}-b_{2}a_{1}} \
{\sf T}(a_{2}b_{2}c_{2}) {\sf T}(a_{1}, b_{1}, c_{1}).
\end{equation}
If we denote the generators of the one parameter subgroups by  $P$ and $Q$
\begin{equation}
P \ = \ {\sf T}(1,0,0) \ \ \ Q \ = \ {\sf T}(0,1,0)
\end{equation}
we obtain the Heisenberg-Weyl commutation relation~\cite{julian}
\begin{equation}
Q P =  \Omega P  Q , \ \ \ \Omega P = P \Omega ,\ \ \
 \Omega Q = Q \Omega
\end{equation}
which together with the periodicity properties
\begin{equation}
Q^{N}  = P^{N} \ = \ \Omega^{N}  =  I
\end{equation}
define the discrete Heisenberg-Weyl group $ HW_{N} $.
The general element~eq.(\ref{HW}) can be written as
\begin{equation}
\label{translations}
{\sf T}(a,b,c) = \Omega^{c} P^{a} Q^{b}
%\left(\begin{array}{ccc}
%1 & a & b \\
%0 & 1 & c \\
%0 & 0 & 1
%\end{array}
%\right)
\end{equation}
The subgroups mentioned previously have the following matrix realizations:
the dilatations:
\begin{equation}
\label{dilatations}
{\sf D}(a,b) = \left(\begin{array}{ccc}
a & 0 & 0 \\
0 & b & 0 \\
0 & 0 & (ab)^{-1}
\end{array}
\right)
\end{equation}
The rotations, which  form the discrete subgroup,
$SO(3,\mathbb{Z}_N)$,  preserves the norm, $(x^1)^2 + (x^2)^2+(x^3)^2$
mod $N$ and are generated by the following matrices:
\begin{equation}
\label{rotations}
\begin{array}{ccc}
\displaystyle
{\sf R}_1(a_1,b_1) =
\left(\begin{array}{ccc}
1 & 0 & 0 \\
0 & a_1 & b_1 \\
0 & -b_1 & a_1
\end{array}
\right), &
\displaystyle
{\sf R}_2(a_2,b_2) =
\left(\begin{array}{ccc}
a_2 & 0 & b_2 \\
0 & 1 & 0 \\
-b_2 & 0 & a_2
\end{array}
\right), &
\displaystyle
{\sf R}_3(a_3,b_3) =
\left(\begin{array}{ccc}
a_3 & b_3 & 0 \\
-b_3 & a_3 & 0 \\
0 & 0 & 1
\end{array}
\right)
\end{array}
\end{equation}
with $a_i^2+b_i^2\equiv\,1\,\mathrm{mod}\,N$, for $i=1,2,3$ through the group
law
$$
{\sf R} ={\sf R}_1(a_1, b_1){\sf R}_2(a_2, b_2){\sf R}_3(a_3, b_3)
$$
Another interesting subgroup of $SL(3,\mathbb{Z}_N)$ is the discrete Lorentz
group, $SO(2,1,\mathbb{Z}_N)$, where, in ${\sf R}_2$ and ${\sf R}_3$ we
replace $-b_2$ and $-b_3$ with $b_2$ and $b_3$ respectively. These elements then preserve the
norm $(x^1)^2-(x^2)^2-(x^3)^2$ mod $N$.

Since $N$ is prime, $\mathbb{Z}_N$ is a finite algebraic field and there exists a
primitive element, $g$, whose successive powers generate all the elements of
the field. If
$N=4k\pm 1$, then the subgroups generated by each ${\sf R}_i$ are cyclic, of
 order $4k$ and they contain the three duality matrices (Fourier transforms)
 for each of the phase space planes (12, 23, 31)\cite{lang}.

To characterize discrete linear Nambu flows
 we must determine the form of the elements of
$SL(3,\mathbb{Z}_N)$ which leave invariant a given vector,
$\bm{a}\in\mathbb{T}_N^3$, i.e. that this be a (left) eigenvector of the
evolution matrix, ${\sf A}$, with eigenvalue unity,
\begin{equation}
\label{left_eigenvector}
\bm{a} = \bm{a}\cdot{\sf A}
\end{equation}
Rotations and translations have, indeed, this property, since they do leave
certain vectors invariant, whereas certain dilatations do not.

For any such vector, $\bm{a}=(a_1,a_2,a_3), a_{i} \in \mathbb{Z}_{N}$,
condition~(\ref{left_eigenvector}) allows us to solve for the elements
$A_{31},A_{32}, A_{33}$, if $ A_{33} $ is different from zero.
The little group of $\bm{a}$ is an $SL(2,\mathbb{Z}_N)$ subgroup. 
Indeed, the evolution
equation, eq.~(\ref{classical_evol}), becomes
\begin{equation}
\label{2Dmapping}
\left[\bm{a}\times\bm{x}\right]_{n+1} = \left[\bm{a}\times\bm{x}\right]_n
\left(
\begin{array}{ccc}
A_{22}-\frac{a_2}{a_3}A_{23} & -(A_{21}-\frac{a_1}{a_3}A_{23}) &
\frac{A_{21}a_2-A_{22}a_1}{a_3}\\
-(A_{12}-\frac{a_2}{a_3}A_{13}) & A_{11}-\frac{a_1}{a_3}A_{13} &
-\frac{A_{11}a_2-A_{12}a_1}{a_3} \\
0 & 0 & 0\\
\end{array}
\right)
\end{equation}
This equation  implies  that the component of any vector $\bm{x}$, 
parallel to $\bm{a}$, is preserved under this evolution, while the components,
that lie on the plane, perpendicular to $\bm{a}$ and are represented by
$\bm{a}\times \bm{x}$, evolve under the $2\times 2$ matrix
\begin{equation}
\label{reduced}
\widetilde{\sf A}\equiv
\left(\begin{array}{cc}
A_{11}-\frac{a_1}{a_3}A_{13} & A_{21}-\frac{a_1}{a_3}A_{23} \\
A_{12}-\frac{a_2}{a_3}A_{13} & A_{22}-\frac{a_2}{a_3}A_{23}
\end{array}
\right)
\end{equation}
(it is noteworthy that $\widetilde{\sf A}$ is the inverse of the $2\times 2$
block in eq.~(\ref{2Dmapping})!)
It is quite straightforward to check that the
determinant of $\widetilde{\sf A}$ is equal to 1 if
$\bm{a}=\bm{a}\cdot{\sf A}$; thus, for, any such ${\sf A}\in
SL(3,\mathbb{Z}_N)$, we have a mapping to an $\widetilde{\sf A}\in
SL(2,\mathbb{Z}_N)$, which is the little subgroup of A which leaves invariant
the vector  $\bm{a}$. This mapping is a group homomorphism, $\widetilde{\sf A}\widetilde{\sf B} = \widetilde{\sf AB}$. 

This  mapping will be useful for the quantization of LNFs in $SL(3,\mathbb{Z}_N)$.

\section{The non-commutative three-torus}\label{nc_3T}

Non-commutative tori play an important role in non-commutative geometry~\cite{rieffel,balachandran}, in M-theory Matrix models\cite{connes} and Quantum Hall effect~\cite{wiegmann,bedos}.
In the present context we need the description of the non-commutative 3-torus, which is appropriate for the study of quantization of linear Nambu flows. 

Let us begin by recalling that it
is possible to embed the Heisenberg-Weyl algebra for one degree of freedom
\begin{equation}
[ x_{1} , x_{2} ] = \mathrm{i} \hbar I
\end{equation}
in the three dimensional noncommutative 3-space
 $\mathbb{R}_\theta^3$
\begin{equation}
 [ x^{i} , x^{j} ] = \mathrm{i} \hbar \epsilon^{ijk} \theta^{k} \ \ \ \
 \ \ \ i,j,k=1,2,3 \ \ \ \ \ \bm{\theta} \in \mathbb{R}^{3}
 \end{equation}
 so that the two-dimensional  quantum phase space is defined by the Casimir \cite{axenides_floratos}
 \begin{equation}
 C \ = \  \bm{\theta} \cdot \bm{x}
 \end{equation}
We can compactify this algebra by considering the algebra of the group elements :
\begin{equation}
T_{\bm{a}} \ = \ e^{\mathrm{i} \bm{a} \cdot \bm{x} } , \ \ \ \bm{a} \in \mathbb{R}^{3}.
\end{equation}
They satisfy 
\begin{equation}
T_{\bm{a}}T_{\bm{b}} \ = \ 
e^{- \frac{\mathrm{i} \hbar}{2} \det (\bm{a},\bm{b},\bm{\theta})} \ \ T_{\bm{a}+\bm{b}},
\end{equation}
These imply
\begin{equation}
[ T_{\bm{a}} , T_{\bm{b}} ] \ = \ 
- 2 \mathrm{i} \sin \left(\frac{\hbar }{2} \det(\bm{a},\bm{b},\bm{\theta} ) \right)\ 
T_{\bm{a}+\bm{b}}.
\end{equation}
 If the quantization and deformation parameters satisfy
\begin{equation}
\hbar \bm{\theta} \ = \frac{2 \pi}{N} ( k_{1}, k_{2}, k_{3} ) \ \ \ \ \ k_{i} \in \mathbb{Z}_{N} \ \ i=1,2,3 ,
\end{equation}
then the Hilbert space of the objects becomes an infinite set of identical copies
of the Hilbert space $ {\mathscr H}_{N} $ of dimension $N$ . 
In this space let us define the discrete
noncommutative 3-torus algebra  as the set generated by three 
$ N \times N $ unitary
matrices $ Q_{i} , i=1,2,3 $ satisfying ( for fixed $k_{i} \mod \ N , 
i=1,2,3 $
\begin{equation}
\label{manin_plane}
Q_{3} Q_{2} = \omega^{k_{1}}  Q_{2} Q_{3}, \ \ \
 Q_{1} Q_{3} = \omega^{k_{2}}  Q_{3} Q_{1} \ \ \
 Q_{2} Q_{1} = \omega^{k_{3}}  Q_{1} Q_{2}.
\end{equation}
with $k_i\in\mathbb{Z}_N$ and $\omega\equiv\exp(2\pi\mathrm{i}/N)$.

The magnetic translation operators can be defined as :
\begin{equation}
J_{\bm{m}} \ = \ \omega^{\frac{1}{2}(k_{3}m_{1}m_{2} + k_{1}m_{2}m_{3}-k_{2}m_{3}m_{1})} \ \
Q_{1}^{m_{1}} Q_{2}^{m_{2}} Q_{3}^{m_{3}}
\end{equation}
The phase is crucial and has to be chosen
so that we have
\begin{equation}
J_{\bm{m}}^\dagger \ = \ J_{-\bm{m}} \ \ \ \ \ \ \ \ \bm{m} \in \mathbb{Z}_{N}^{3}.
\end{equation}
We then also find that 
\begin{equation}
J_{\bm{m}} J_{\bm{n}} \ = \ \omega^{-\frac{1}{2} \det(\bm{k},\bm{m},\bm{n})} \ J_{\bm{m}+\bm{n}}, \ \ \ \ \ \ \ \bm{m},\bm{n} \in \mathbb{Z}_{N}^{3} ,
\end{equation}
We can see that these commutation relations imply the existence of central element of the algebra
$ C_{T} = Q_{1}^{k_{1}}Q_{2}^{k_{2}} Q_{3}^{k_{3}}$ . If the representation is
irreducible, then $C$ must be proportional to the unit element, up to a phase,
$C_T  = \omega^{c_N}\times  I $, with $c_N\in\mathbb{Z}_N$. Consider now the elements
$J_{\alpha\bm{k}}$, with $\alpha\in\mathbb{Z}_N$. It is easy to check that
they are pure phases
\begin{equation}
\label{phasesJ}
J_{\alpha\bm{k}} = \omega^{ \frac{\alpha^2}{2}k_1k_2k_3 + \alpha (c_N-k_1 k_2 k_3)}\times I 
\end{equation} 
Therefore the $N^3$ magnetic translations are divided into a subgroup of $N$
phases and a set of $N^2$ unitary matrices $J_{\bm{m}}$, where $\bm{m}$ is
orthogonal to $\bm{k}$. This structure resembles that of the discrete
Heisenberg--Weyl group, $HW_N$. The magnetic translation operators thus depend
on the vector $\bm{k}$ and we shall henceforth explicitly highlight this by
writing them as $J_{\bm{m}}(\bm{k})$. The commutation relations between 
$J_{\bm{m}}(\bm{k})$ and $J_{\bm{m}}(\bm{k}')$ can be computed once we shall
establish the relation between these magnetic translatons and the
Heisenberg--Weyl generators.

The classical action of the discrete map ${\sf A} \in SL(3,\mathbb{Z}_{N}) $
  on the points $ \bm{m} = ( m_{1} , m_{2}, m_{3}) $ of the torus $ T_{N}^{3} $ was reduced in the previous section to the action of
$ \widetilde{\sf A} \in SL(2, \mathbb{Z}_{N} )$ on the points
$ \widetilde{\bm{m}}\equiv (a_{3}m_{1}-a_{1}m_{3},a_{3}m_{2}-a_{2}m_{3})$
of the plane
  orthogonal to $\bm{a}$. 

If we restrict $\bm{m}$ to this plane, we obtain 
\begin{equation}
\widetilde{\bm{m}} = (m_1,m_2)T(\bm{a})\equiv 
(m_1,m_2)\left(\begin{array}{cc} 
\displaystyle
\frac{1-a_2^2}{a_3} & \displaystyle \frac{a_1a_2}{a_3} \\
\displaystyle
\frac{a_1a_2}{a_3} & 
\displaystyle
\frac{1-a_1^2}{a_3}\end{array}\right)
\end{equation}
where we have assumed that $a_1^2 + a_2^2 + a_3^2\equiv 1\,\mathrm{mod}\,N$.
In this case $T(\bm{a})$ is an element of $SL(2,\mathbb{Z}_N)$.

Upon quantization on this discrete two--dimensional phase space, 
we should employ  the discrete Heisenberg-Weyl
group, generated by the clock and shift $ N \times N $ matrices $Q, P$ \cite{afn}

\begin{equation}
Q_{k,l} \ = \ \omega^{k} \delta_{k,l},  \ \ \ \  P_{k,l} = \delta_{k,l+1}, \  \ \ \ k,l\in \mathbb{Z}_{N}
\end{equation}
which satisfy
\begin{equation}
QP\ = \omega P Q
\end{equation}
The corresponding  {\em two-dimensional} magnetic translations defined by 
\begin{equation}
J_{r,s} \ = \ \omega^{\frac{rs}{2}} P^{r}Q^{s} \ \ \ \ \ \ \ \ \ \ r,s \in \mathbb{Z}_{N}
\end{equation}

satisfy the relations
\begin{equation}
\begin{array}{l}
\displaystyle
J_{r,s} J_{r',s'} \ = \ \omega^{\frac{r' s - r s'}{2}} J_{r+r',s+s'}\\
\\
\displaystyle 
\left[J_{r,s}\right]^{\dagger} \ = \ J_{-r, -s} \\
\end{array}
\end{equation}
where $r,s,r',s'\in \mathbb{Z}_N$.

We now identify the points of the torus $\mathbb{T}_N^2$ on which the map
$\widetilde{\sf A}$ acts with the indices, $(r,s)$ 
 of the two-dimensional magnetic
translations, $J_{r,s}$ through
\begin{equation}
r\equiv a_{3}m_{1}-a_{1}m_{3} \ \ \ \ \ s\equiv a_{3}m_{2}-a_{2}m_{3}
\end{equation}
as 
\begin{equation}
J_{\bm{m}}(\bm{k}) \ = \ J_{(m_1,m_2)T(\bm{a})}
\end{equation}
This {\em Ansatz}  implies the relations,
\begin{eqnarray}
\label{embedding}
J_{1,0,0} &=& Q_{1} = J_{a_{3},0} = P^{a_{3}} \ \ \ \nonumber   \\
J_{0,1,0} &=& Q_{2} = J_{0 , a_{3}} = Q^{a_{3}} \ \ \ \nonumber   \\
J_{0,0,1} &=& Q_{3} = J_{-a_{1}, -a_{2}} = \omega^{\frac{a_{1}a_{2}}{2}} P^{-a_{1}} Q^{-a_{2}}
\end{eqnarray}
and from the commutation relations  of the operators 
$ Q_{1} , Q_{2}, Q_{3} $ we find
\begin{equation}
k_{1}=a_{1}a_{3} \ \ \ , \ \ \ k_{2}=a_{2}a_{3} \ \ \  , \ \ \ k_{3}=a_{3}^{2}
\end{equation}

This identification also fixes the phase, $c_N$ of the Casimir in
eq.~(\ref{phasesJ}) as 
\begin{equation}
\label{casimir_phase}
c_N = \frac{k_1 k_2 k_3}{2} = \frac{a_1 a_2 a_3^4}{2}
\end{equation}
Thus the $N$ phases have been eliminated and only magnetic translations in the
plane orthogonal to the vector $\bm{a}$ survive. 
It is possible to represent 
the algebra of eq.~(\ref{manin_plane}) by $3 \times 3 $ matrices of $SL(3,\mathbb{Z}_{N})$ substituting in
eq.~(\ref{embedding}) the $ N \times N $ matrices $P , Q ,  \omega \cdot I $
by the matrices $P={\sf T}(1,0,0),Q={\sf T}(0,1,0),\Omega={\sf T}(0,0,1)$
of the $3 \times 3 $ Heisenberg-Weyl group in eq.(\ref{translations}). 

In order to generate the full magnetic translation group of the three
 dimensional, discrete torus, we must consider three, mutually orthogonal,
 planes and their corresponding $J_{\bm{m}}(\bm{a})$'s. For example,
 $\bm{a}=(1,0,0)$, $\bm{a}=(0,1,0)$ and $\bm{a} = (0,0,1)$. Starting from the
 $1-2$ plane and applying discrete rotations of $SO(3,\mathbb{Z}_N)$,
 cf. eq.~(\ref{rotations}), we can generate the other two. To construct the
 corresponding $J_{\bm{m}}$'s for the $2-3$ and $3-1$ planes, we must construct 
the  corresponding unitary, $N\times N$ operators, $U({\sf R}_{1,2,3})$. 
This remains to be done.
 
In the next section we shall apply the above results  for the
quantization of the classical Nambu mechanics in  the case of linear
flows, for a fixed plane. 

\section{Quantization of Linear Nambu Flows on a Discretized 3-Torus}\label{Quant3T}
There is a long-standing problem on how to quantize Nambu mechanics and there
are various proposals, which, however, do not respect the fundamental
properties of the classical Nambu bracket, such as Leibniz and the fundamental
identity~\cite{nambu, tak, minic, yoneya, matsuo_ho,
  kawamura}. 

Quantization of the classical dynamics, $\bm{x}_{n+1} = \bm{x}_n{\sf A}$, for
${\sf A}\in SL(3,\mathbb{Z}_N)$, means constructing a unitary operator,
$U({\sf A})$ as a $N\times N$ unitary matrix, that satisfies
\begin{equation}
\label{metaplectic_3D}
U^\dagger({\sf A}) J_{\bm{m}}U({\sf A}) = J_{\bm{m}\cdot{\sf A}}
\end{equation}
in the basis of the complete set of
three-dimensional magnetic translations of the
non-commutative three-torus. This would realize
 the $N$-dimensional {\em metaplectic} representation of
of the double cover of $SL(3,\mathbb{Z}_N)$.
For rigorous mathematical results pertaining to the metaplectic representation
of the double cover of $SL(3,\mathbb{F})$, for $\mathbb{F}$ a local field we
refer to the literature~\cite{kazhdan}; for $\mathbb{F}=\mathbb{R}$
cf.~\cite{torasso}.

The results of the previous sections allow us to do this, for the case of
linear Nambu flows. Indeed, we found that, for those flows, the classical
evolution equation, $\bm{x}_{n+1} = \bm{x}_n\cdot {\sf A}$ may be written
inthe form of
$$
[\bm{a}\times\bm{x}_{n+1}] =
[\bm{a}\times\bm{x}_n]\left(
\begin{array}{cc}
\widetilde{\sf A}^{-1} & \widetilde{\bm{a}} \\
\bm{0}        & 0
\end{array}
\right)
$$
where $\widetilde{\sf A}$ is given by eq.~(\ref{reduced}) and the vector
$\widetilde{\bm{a}}^T\equiv
\left(
(A_{21}a_2 - A_{22}a_1)/a_3, -(A_{11}a_2-A_{12}a_1)/a_3
\right)
$
and thus the interesting dynamical variables are the combinations
$[\bm{a}\times\bm{x}_n]_1\equiv a_2 x_3-a_3x_2$ and
$[\bm{a}\times\bm{x}_n]_2\equiv a_3 x_1-a_1x_3$; the other component may be
expressed as a linear combination of these. Since $\widetilde{\sf A}\in
SL(2,\mathbb{Z}_N)$ when ${\sf A}\in SL(3,\mathbb{Z}_N)$ and
$\bm{a}=\bm{a}\cdot{\sf A}$, we know how to construct the unitary operator,
$U(\widetilde{\sf A})$,
that realizes the metaplectic representation for $\widetilde{\sf A}$.
Furthermore, we may verify that $\widetilde{\sf A}\cdot\widetilde{\sf B} =
\widetilde{\sf AB}$, for any two matrices
 ${\sf A}, {\sf B}\in SL(3,\mathbb{Z}_N)$
that have $\bm{a}$ as a common eigenvector, with eigenvalue unity,
$\bm{a}=\bm{a}\cdot{\sf A}$, $\bm{a}=\bm{a}\cdot{\sf B}$.
So we can write the following, commuting, diagram
\begin{equation}
\label{commuting_diag}
\begin{array}{ccc}
{\sf A}         & \longrightarrow & \widetilde{\sf A} \\
\downarrow   &                & \downarrow    \\
U({\sf A})       &  \longrightarrow              &  U(\widetilde{\sf A})\\
\end{array}
\end{equation}
To construct the corresponding (unitary) evolution operator, $U({\sf A})$,
we thus use the metaplectic representation of $SL(2,\mathbb{Z}_N)$ for the
``reduced'' $2\times 2$ matrix, $\widetilde{\sf A}$ of eq.~(\ref{reduced})
which satisfies
\begin{equation}
\label{reduced_metapl}
U({\sf\widetilde{A}})^\dagger J_{r,s} U({\sf\widetilde{A}}) =
J_{(r,s){\sf\widetilde{A}}}
\end{equation}
and is given, for any element
$$
\widetilde{ {\sf A} } \equiv \left(\begin{array}{cc} a & b \\ c & d \\
\end{array}\right)\in SL(2,\mathbb{Z}_N)
$$
by the expression~\cite{athanasiu_floratos,afn}
\begin{equation}
\label{UAexplicit}
U(\widetilde{{\sf A}})_{k,l} =
\frac{\sigma_N(c)}{\sqrt{N}}\omega^{\frac{ak^2-2kl+dl^2}{2c}}
\end{equation}
where
$$
\sigma_N(c)\equiv\frac{1}{\sqrt{N}}\sum_{r=0}^{N-1}\omega^{c\cdot r^2}
$$
is the Gauss sum~\cite{lang}.

The prefactor, $\sigma_N(c)$ ensures that this representation is not only
projective, but {\em faithful}, i.e., for {\em any} two matrices, 
$\widetilde{A}$ and $\widetilde{B}$, elements of $SL(2,\mathbb{Z}_N)$, we have
that $U(\widetilde{\sf AB}) = U(\widetilde{A})U(\widetilde{B})$.

Having thus shown that, for linear Nambu flows, the interesting dynamics takes
place on a plane perpendicular to the vector $\bm{a}$, that enters in the
definition of $H_2\equiv \bm{a}\cdot\bm{x}$, we can understand why the
construction of the unitary operator $U({\sf A})\equiv U(\widetilde{\sf A})$
amounts to a quantization of discrete position and momenta: 
From the classical vectors $\bm{m}$ and $\bm{a}$
we construct the corresponding position and momentum variables $r\equiv
a_3 m_1-a_1 m_3$ and $s\equiv a_3 m_2-a_2 m_3$ respectively. These evolve
using the operator $\widetilde{\sf A}$, while the $N-$dimensional,complex
vector (wavefunction),in the position representation depends on
$r=0,1,2,\ldots,N-1$ and evolves according to $U(\widetilde{\sf A})$. From
this operator, we may calculate the average value(s) of physical
observables,as well as correlation functions of the flow,using standard
quantum mechanical techniques.For physically interesting subgroups of
$SL(3,\mathbb{Z}_N)$, mentioned in section~\ref{discrete_phsp},we may find the
eigenstates and eigenvalues of $U(\widetilde{\sf A})$ explicitly. This will be
reported elsewhere. 

\section{Conclusions}\label{Concl}
We have constructed the quantization of Nambu mechanics ,for the case of linear flows on the 
discrete 3-dimensional torus considered as a phase space.Our method proposes also a scheme for the quantization of the Nambu 3-bracket as the algebra of the foliation of the non-commutative 3-torus by a family
of Heisenberg-Weyl groups of all of its linear 2-dimensional subspaces (non-commutative 2-tori).
 The key idea was to use the metaplectic representation of 
$SL(3, \mathbb{Z}_N)$, induced by that of $SL(2,\mathbb{Z}_N)$.
Considering potential applications ,our method can be extended to the full set of 
discrete linear flows ,not necessarily of the Nambu type (not having invariant 2-dimensional subspaces).
This will lead to the quantization of strong arithmetic chaos~\cite{vivaldi} 
on the discrete
3-torus and can be used as a toy model for quantization of turbulent maps~\cite{lima}.
Considering the coset $SL(3,\mathbb{Z}_N)/SO(3,\mathbb{Z}_N)$ we can construct the corresponding quantum coherent
states( discrete orthogonal wavelets)
for the wavelet transform of three dimensional objects(quantum
tomography 
 in the spirit of~\cite{athanasiu_floratos,coen}.
 Another possible direction is the study of discrete non-commutative solitons in 3-dimensions
 using non-dispersive t'Hooft states~\cite{flonic2002}..
 Concluding we believe that the proposed framework of quantization for Nambu mechanics will lead to new insights
 for the quantization of volume preserving diffeomorphism group in 3-dimensions.


\newpage

\begin{thebibliography}{99}

\bibitem{bagger_lambert}J.~Bagger and N.~Lambert,
  %``Modeling multiple M2's,''
  {\sl Phys.\ Rev.}\   {\bf D75} (2007) 045020
  [arXiv:hep-th/0611108].
  %%CITATION = PHRVA,D75,045020;%%

\bibitem{nambu} Y. Nambu,
%``Generalized Hamiltonian dynamics,''
  {\sl Phys.\ Rev.}\   {\bf D7} (1973) 2405.
  %%CITATION = PHRVA,D7,2405;%%
\bibitem{tak}
L.~Takhtajan,
  %``On Foundation Of The Generalized Nambu Mechanics (Second Version),''
  {\sl Commun.\ Math.\ Phys.}\  {\bf 160} (1994) 295
  [arXiv:hep-th/9301111].
  %%CITATION = CMPHA,160,295;%%
\bibitem{hop}
J.~Hoppe,
  %``On M-Algebras, the Quantisation of Nambu-Mechanics, and Volume Preserving
  %Diffeomorphisms,''
  {\sl Helv.\ Phys.\ Acta}  {\bf 70} (1997) 302
  [arXiv:hep-th/9602020].
\bibitem{yoneya}H.~Awata, M.~Li, D.~Minic and T.~Yoneya,
  %``On the quantization of Nambu brackets,''
  {\sl JHEP} {\bf 0102} (2001) 013
  [arXiv:hep-th/9906248].
  %%CITATION = JHEPA,0102,013;%%
 \bibitem{minic}D.~Minic and H.~C.~Tze,  %``Nambu quantum mechanics: A nonlinear generalization of geometric  quantum
  %mechanics,''
  {\sl Phys.\ Lett.}\   {\bf B536} (2002) 305
  [arXiv:hep-th/0202173].
  %%CITATION = PHLTA,B536,305;%%
\bibitem{kawamura}Y.~Kawamura,
  %``Cubic matrix, Nambu mechanics and beyond,''
  {\sl Prog.\ Theor.\ Phys.}\  {\bf 109} (2003) 153
  [arXiv:hep-th/0207054].
  %%CITATION = PTPKA,109,153;%%
\bibitem{zachos}
 C.~K.~Zachos and T.~Curtright,
  %``Branes, quantum Nambu brackets, and the hydrogen atom,''
  {\sl Czech.\ J.\ Phys.}\  {\bf 54} (2004) 1393
  [arXiv:math-ph/0408012].
  %%CITATION = CZYPA,54,1393;%%
\bibitem{gustavsson}A.~Gustavsson,
{\em Algebraic structures on parallel M2-branes},
  arXiv:0709.1260 [hep-th].
  %%CITATION = ARXIV:0709.1260;%%
\bibitem{papadopoulos}G.~Papadopoulos,
  %``M2-branes, 3-Lie Algebras and Plucker relations,''
  {\sl JHEP} {\bf 0805} (2008) 054
  [arXiv:0804.2662 [hep-th]].
  %%CITATION = JHEPA,0805,054;%%
\bibitem{matsuo_ho}P.~M.~Ho, R.~C.~Hou and Y.~Matsuo,
  {\sl JHEP}  {\bf 0806} (2008) 020
  [arXiv:0804.2110 [hep-th]].
  %%CITATION = ARXIV:0804.2110;%%
\bibitem{matsuo_ho2}P.~M.~Ho and Y.~Matsuo,
%  {\em M5 from M2},
{\sl JHEP} {\bf 0806} (2008) 105
  [arXiv:0804.3629 [hep-th]].
  %%CITATION = ARXIV:0804.3629;%%
P.~M.~Ho, Y.~Imamura and Y.~Matsuo,
%  {\em M2 to D2 revisited},
{\sl JHEP} {\bf 0807} (2008) 003,
  [arXiv:0805.1202 [hep-th]].
  %%CITATION = ARXIV:0805.1202;%%
\bibitem{axenides_floratos} M. Axenides and
  E. G. Floratos, {\em Nambu-Lie 3-Algebras on Fuzzy 3-Manifolds},, [arXiv:0809.3493[hep-th]]%%CITATION  = ARXIV:0809.3493 
\bibitem{marsden_weinstein} J. Marsden and A. Weinstein, {\sl Physica} {\bf
  7D} (1983) 305.
\bibitem{lamb}
H.~Lamb , {\em Hydrodynamics}, (Cambridge University Press, UK , 1932) , p.248.
\bibitem{Moff}
K.~Bajer and H.~K.~Bajer and H.~K.~Moffatt ,
%'' On a class of Steady Confined Stokes Flows with Chaotic Streamlines ''
{\sl J.Fluid Mech.}  { \bf 212 } (1990) 337.
\bibitem{arnold}
V.I. Arnold and B. A. Khesin, {\em Topological Methods in Hydrodynamics} 
(Springer-Verlag) (1998).
\bibitem{rajeev}
S.~G.~Rajeev ,
% '' Incompressible Fluids '',
{\sl Int. J.Mod. .Phys.}  { \bf A20 } (2005) 6122. %%CITATION = IMPAE,A20,6122;%%

\bibitem{julian} J.Schwinger, 
 {\em Quantum Kinematics and Dynamics}, Benjamin (1970).
\bibitem{lang}
S. L. Lang, {\em Analytic Number Theory}, Addison-Wesley (1970).
\bibitem{rieffel} M. Rieffel , {\sl Can. J. Math.} {\bf XL} (1988) 257; Yu. Manin, 
{\em Theta Functions, Quantum Tori and Heisenberg Groups}, [arXiv:math.AG/0011197]
%%CITATION = MATH.AG 0011197
; A. Schwarz, {\em Theta Functions on Non-commutative Tori}, [arXiv:math/0107186].
%%CITATION =  MATH 0107186
\bibitem{vivaldi} I.C. Percival and F. Vivaldi,{\sl Physica} {\bf D25} (1987) 105.
\bibitem{lima} E. Ugalde, R. Lima, {\sl Physica} {\bf D95} (1996) 144.
\bibitem{balachandran} A. P. Balachandran, S. Kurkcuoglu and S. Vaidya, {\em Lectures on Fuzzy and Fuzzy SUSY Physics}, World Scientific (2007).
\bibitem{connes} A.~Connes, M.~R.~Douglas and A.~S.~Schwarz,
  %%``Noncommutative geometry and matrix theory: Compactification on tori,''
  {\sl JHEP}  {\bf 9802} (1998) 003
  [arXiv:hep-th/9711162].
  %%CITATION = JHEPA,9802,003;%%
 \bibitem{wiegmann} D. Khveshchenko and P.Wiegmann, {\sl Phys. Lett.} {\bf B225} (1989) 279
 %%CITATION = PHLTA,B225,279
 \bibitem{bedos} E. Bedos, {\sl J. Geom. Phys.} {\bf 30} (1999) 204
\bibitem{kazhdan} Y. Flicker, D. Kazhdan and G.Savin, {\em Explicit Realization
  of a Metaplectic Representation}, {\sl J. Analyse Math.} {\bf 55} (1990)
  17.
\bibitem{torasso} P.Torasso, {\sl Acta Math.} {\bf 150} (1983) 153.
\bibitem{athanasiu_floratos} G. G. Athanasiu and  E.G. Floratos, 
{\sl Phys. Lett.} {\bf B352} (1995) 105; %%CITATION = PHLTA B352,105%% 
G. G. Athanasiu and E. G. Floratos,
{\sl Nucl. Phys.} {\bf B425} (1994) 343;
  %%CITATION = NUPHA,B425,343;%
E. G. Floratos and S. Nicolis, {\sl J. Phys. A} {\bf 31} (1998) 3961
[arXiv:hep-th/9508111]
%%CITATION = JPAGB,A31,3961%%
\bibitem{afn} G. G. Athanasiu, E. G. Floratos and S. Nicolis,
  %``Holomorphic Quantization on the Torus and Finite Quantum Mechanics,''
{\sl   J.\ Phys.\ A } {\bf 29} (1996) 6737
  [arXiv:hep-th/9509098].
  %%CITATION = JPAGB,A29,6737;%%
%``Fast Quantum Maps,''
 {\sl  J.\ Phys.\ A}  {\bf 31} (1998) L655
  [arXiv:math-ph/9805012].
  %%CITATION = JPAGB,A31,L655;%%
\bibitem{coen} S. Coen, {\sl Phys. Rev. Lett.} {\bf 83} (1999) 2494.
\bibitem{flonic2002}
G.~'t Hooft,
  %``QUANTIZATION OF DISCRETE DETERMINISTIC THEORIES BY HILBERT SPACE
  %EXTENSION,''
  {\sl Nucl.\ Phys.}   {\bf B342} (1990) 471.
  %%CITATION = NUPHA,B342,471;%%
E.G. Floratos and G. Leontaris,
{\sl Phys.\ Lett.}  {\bf B412} (1997) 35
  [arXiv:hep-th/9706156];  %%CITATION = PHLTA,B412,35%%
E. G. Floratos and S. Nicolis, {\em Non-Commutative Solitons
  in Finite Quantum Mechanics}, {\sl Nucl. Phys. Proc. Suppl.} {\bf 119}
  (2002) 947, [arXiv:hep-lat/0209032] %%CITATION = NUPHZ,119,947;%%
\end{thebibliography}
\end{document}